\documentclass[conference]{IEEEtran}
\IEEEoverridecommandlockouts
\usepackage{cite}
\usepackage{amsmath,amssymb,amsfonts}
\usepackage{algorithmic}
\usepackage{graphicx}
\usepackage{textcomp}
\usepackage{xcolor}
\usepackage{url}
\usepackage{comment}

\usepackage{multicol} 
\usepackage{longtable}
\usepackage{multirow}
\usepackage{float}
\usepackage{graphicx,pifont}
\usepackage{booktabs}
\usepackage{dirtytalk} 
\usepackage{placeins} 
\usepackage{rotating}
\usepackage{lscape} 
\usepackage[normalem]{ulem}
\usepackage{caption} 
\usepackage{float} 
\usepackage{csquotes} 
\usepackage{latexsym}
\usepackage[most]{tcolorbox}
\usepackage{subfig} 
\usepackage{float} 
\def\BibTeX{{\rm B\kern-.05em{\sc i\kern-.025em b}\kern-.08em
    T\kern-.1667em\lower.7ex\hbox{E}\kern-.125emX}}
\begin{document}


\title{Personalized Guidelines for Design, Implementation and Evaluation of Anti-phishing Interventions\\}

 \makeatletter
 \newcommand{\linebreakand}{%
  \end{@IEEEauthorhalign}
   \hfill\mbox{}\par
   \mbox{}\hfill\begin{@IEEEauthorhalign}
 }
 \makeatother
\author{\IEEEauthorblockN{Orvila Sarker}
\IEEEauthorblockA{
\textit{University of Adelaide}\\
Cyber Security Cooperative Research Centre \\
orvila.sarker@adelaide.edu.au}
\and
\IEEEauthorblockN{Sherif Haggag}
\IEEEauthorblockA{
\textit{University of Adelaide} \\
sherif.haggag@adelaide.edu.au}
\linebreakand 

\IEEEauthorblockN{Asangi Jayatilaka}
\IEEEauthorblockA{
\textit{University of Adelaide}\\
Cyber Security Cooperative Research Centre\\%
asangi.jayatilaka@adelaide.edu.au}
\and
\IEEEauthorblockN{Chelsea Liu}
\IEEEauthorblockA{
\textit{University of Adelaide} \\
chelsea.liu@adelaide.edu.au}

}

\maketitle

\IEEEpubid{978-1-6654-5223-6/23\$31.00 \copyright\ 2023 Crown}
\enlargethispage{-\baselineskip}

\begin{abstract}
Background: Current anti-phishing interventions, which typically involve one-size-fits-all solutions, suffer from limitations such as inadequate usability and poor implementation. Human-centric challenges in anti-phishing technologies remain little understood. Research shows a deficiency in the comprehension of end-user preferences, mental states, and cognitive requirements by developers and practitioners involved in the design, implementation, and evaluation of anti-phishing interventions. Aims: This study addresses the current lack of resources and guidelines for the design, implementation and evaluation of anti-phishing interventions, by presenting personalized guidelines to the developers and practitioners. Method: Through an analysis of 53 academic studies and 16 items of grey literature studies, we systematically identified the challenges and recommendations within the anti-phishing interventions, across different practitioner groups and intervention types. Results: We identified 22 dominant factors at the individual, technical, and organizational levels, that affected the effectiveness of anti-phishing interventions and, accordingly, reported 41 guidelines based on the suggestions and recommendations provided in the studies to improve the outcome of anti-phishing interventions. Conclusions: Our dominant factors can help developers and practitioners enhance their understanding of human-centric, technical and organizational issues in anti-phishing interventions. Our customized guidelines can empower developers and practitioners to counteract phishing attacks.
\end{abstract}

\section{Introduction}
Phishing is a form of cyber attacks committed by using fraudulent and deceitful communication techniques, such as emails or messages, to entrap users into providing sensitive personal information, such as passwords, credit card details, or social security numbers \cite{jenkins2022phished}. Automated detection techniques can provide a line of defence against phishing (e.g., \cite{alani2022phishnot}), however, ultimately end users serve as the final safeguard. The abundance of successful recent phishing attacks indicates that further efforts are needed to enhance end users' phishing education, training, and awareness (PETA) \cite{wash2021knowledge,APWG2022}. 

Empirical investigations have demonstrated that tailored design, implementation, and evaluation of phishing interventions can be efficacious in helping users recognize and mitigate phishing hazards \cite{kaiser2021adapting, arachchilage2016phishing, sheng2007anti, kumaraguru2009school}. A phishing intervention refers to any anti-phishing system, software, tool, or framework that helps users deal with a phishing attack and requires user intervention \cite{franz2021sok}. The cognitive needs and mental status of individual end users play an important role in determining the effectiveness of phishing intervention and, consequently, ought to take into consideration by the developers and practitioners during the design, implementation and evaluation process of phishing intervention \cite{li2014towards}. However, evidence shows that developers often neglect end-users' decision-making processes and cognitive limitations in the design, implementation, and evaluation of phishing interventions. This leads to human-centric weaknesses or usability issues, rendering the end-users susceptible to phishing attacks \cite{egelman2008you, wu2006security, alsharnouby2015phishing}. In order to ensure the efficacy and usefulness of anti-phishing interventions, their key features such as the content and methods of anti-phishing intervention must be tailored to the needs of individual users\cite{reuter2022quarter,salamah2022importance,bullee2020effective}.
\enlargethispage{-\baselineskip}

\textbf{Motivation:}
The effectiveness of phishing education, training, and awareness (PETA) interventions  significantly depends on decisions made by various developers and practitioners involved in the design, implementation, and evaluation of these interventions. Studies document numerous examples of failures in such decision-making: some email providers do not use Simple Mail Transfer Protocol (SMTP) authentication mechanisms, thus allowing the attackers to send emails from spoofed email addresses \cite{hu2018end}; many web developers of e-commerce enterprises fail to employ Secure Socket Layer (SSL) to secure their login page \cite{wu2006security}; organizations' security officers often conduct phishing training without following any formal policies \cite{althobaiti2021case} and use unrealistic or irrelevant email templates \cite{egelman2008you,caputo2013going,burns2019spear}; developers design interventions with complex user interfaces that require specialized knowledge to install and utilize \cite{cj2018phishy,wen2019hack,marforio2016evaluation,althobaiti2021don}. These examples highlight the need for a structured set of guidelines provided to developers and practitioners to help them comprehend the diverse range of needs and challenges faced by end-users. However, currently, there is a limited availability of such guidelines for developers and practitioners. Moreover, most existing resources are intended for a singular group of practitioners or a particular class of cyber security interventions but are not specific to phishing interventions. For instance, Lujo et al. \cite{bauer2013warning} and Lynsay et al. \cite{shepherd2018design} reported guidelines for browser security warning design (this is a type of phishing intervention), and Mirium et al. \cite{galikova2021toward} provided guidelines for the development of cyber security games (not directly related to phishing). Consequently, the target user group of these guidelines are primarily designers or developers of phishing interventions, with limited relevance to other practitioners such as organization managers or  cyber security officers. Guidelines for IT security management (e.g., proposed by Pooya et al. \cite{jaferian2008guidelines} and Sonia et al. \cite{chiasson2007even}) do not offer insights specific to phishing prevention. Overall, current guidelines and best practices for anti-phishing interventions (particularly on their design, implementation, and evaluation) are scattered around various academic and grey literature studies and not presented to practitioners in an easily accessible and personalized format. 

In light of these needs and challenges faced by developers and cyber security practitioners, we aim to investigate the following research question in our study: \textit{what guidance can be provided to support the developers and practitioners in addressing usability issues in anti-phishing interventions from an individual, technical and organizational perspective?}
Our study has devised a set of guidelines by synthesizing the existing literature to aid practitioners to obtain a more holistic understanding of end-user needs and experiences, in order to enhance the effectiveness of the design, implementation, and evaluation of anti-phishing interventions.

\textbf{Contributions:} 
\textbullet\ From the existing literature, we have identified \textit{22 dominant factors} which impact the effectiveness of anti-phishing interventions. These include 15 human factors (such as age, complacency, and educational qualification), 4 technical factors (such as device type and gamer type) and 3 organizational factors (such as organizational position and working hours). Our presented dominant factors can assist practitioners to attain a deeper understanding of the important determinants of the success of phishing interventions. \textbullet\ We have reported \textit{41 guidelines} on the design, implementation, and evaluation of anti-phishing interventions, which are systematically compiled based on recommendations derived from a comprehensive Multi-vocal Literature Review (MLR) of 69 studies, thus making the guidelines the first of its kind in terms of its comprehensiveness and breadth of coverage. Our devised guidelines can be a useful resource to aid practitioners to improve the efficacy of the design, implementation and evaluation of anti-phishing interventions.
 
\begin{figure}[t] 
\centering
\includegraphics[trim=32 55 85 18,clip, scale = 0.3]{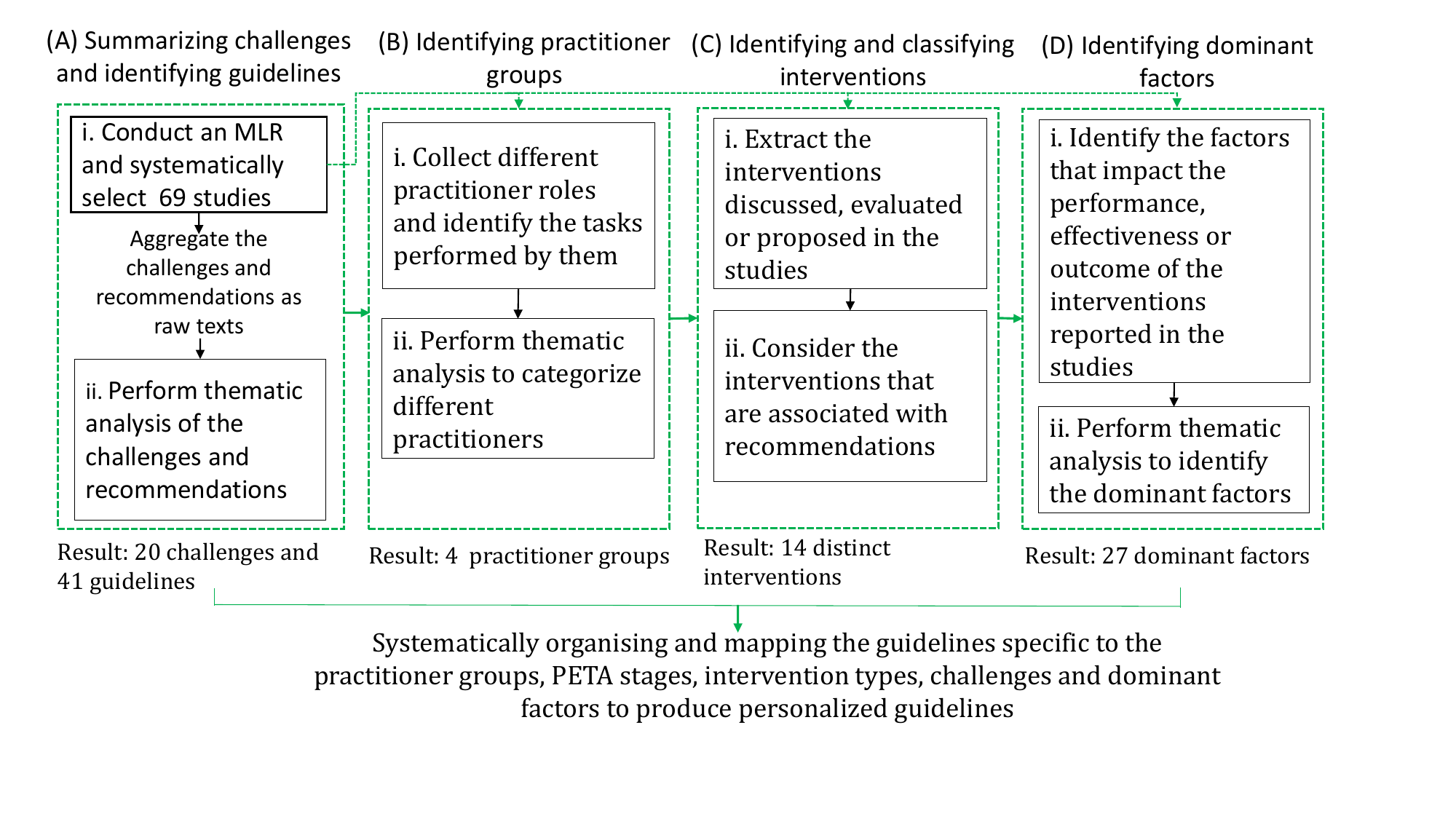}
 \caption{Methodology of this study}
 \label{Methodology of this study}
 \end{figure}
\section{Methodology} 
Our methodology, summarized in Fig \ref{Methodology of this study}, consists of five steps, as detailed in this Section.

\subsection{Summarizing challenges and identifying guidelines}
\label{section:Summarising challenges and developing guidelines}

\subsubsection{Conduct an MLR and systematically select 69 studies}

We conduct an MLR with the aim of identifying the challenges and recommendations concerning anti-phishing interventions. As anti-phishing interventions are inherently industry-oriented, we incorporate the important perspectives of practitioners by including grey literature studies. Our MLR adheres to the protocols outlined by Kitchenham and Charters \cite{kitchenham2007guidelines} and Garousi et al. (2019) \cite{garousi2019guidelines}. 

In our study selection process, we utilized the Scopus \cite{scopus} database, as it offers greater breadth and depth of academic literature compared to other databases \cite{zahedi2016systematic,shahin2020architectural}. We ran a pilot study with ACM digital library and IEEE Xplore digital library to ensure that Scopus is comprehensive. Following a thorough examination of the search keywords used in prior review papers in this field (e.g., \cite{ desolda2021human, baki2021sixteen}) and refinement through conducting pilot searches, we use the search keywords \say{aware*} or \say{interven*} or \say{nudge*} or \say{warn*} or \say{protect*} or \say{security indicators} or \say{alert*} along with the keyword \say{phish*} to collect the academic studies. To ensure the quality of the collected studies, we only included studies that had a CORE \cite{coredatabase} rank of A*, A and B. The CORE ranking system aims to ensure high-quality standards and rigorous peer review processes for selected journals and conferences \cite{core}. We omitted papers that had a CORE rank B and were published before 2012. The reason for this is that our pilot study revealed that a number of CORE B papers that were published prior to 2012 proposed client-side anti-phishing tools without conducting a real-world evaluation of their usability (e.g., Passpet \cite{yee2006passpet}). We excluded studies that have provided automated anti-phishing solutions (defined in \cite{jampen2020don}) and research that were not written in English, short papers (less than 6 pages), and literature survey papers. 

To collect the grey literature studies we employed Google as our search database. Google serves as a widely accepted and utilized search engine for gathering grey literature study \cite{garousi2017software, islam2019multi, butijn2020blockchains}. For the grey literature study collection, we used the search terms \say{education}, \say{training}, \say{awareness} with the term \say{phish*}. The rationale for employing an alternative search query distinct from the one employed in the academic study stems from the search methodology employed by Google; Google conducts searches using the specified search terms across its entire index of webpages \cite{cascavilla2021cybercrime}. Hence, apart from the duplicated webpages and academic studies that have already been identified in Scopus, our pilot study using the identical search string as employed in academic studies yielded numerous extraneous results. Therefore, we used different search terms to obtain more relevant results following the suggestions by Garousi et. al \cite{garousi2019guidelines}.

The grey literature study collection process concluded on Google's 16th page as no new or redundant information was identified, as recommended by Garousi et al \cite{garousi2019guidelines}. To ensure the quality of the grey literature studies, we assessed the publication's authority,  methodology, presence of reliable references, the date of publication, the novelty of the article, and the article's outlet type as suggested by Garousi et al \cite{garousi2019guidelines}. 
More information on our literature selection process, including the search string, inclusion/exclusion criteria, quality assessment criteria, and the list of selected studies in the academic and grey literature, is provided in our online appendix \cite{supplementary}.

After applying the inclusion/exclusion criteria and data quality assessment, we collected a total of 69 studies, including 53 academic and 16 grey literature studies. We use the symbol P[*] to denote the studies throughout the rest of the paper.

\subsubsection{Perform thematic
analysis of the
challenges and recommendations} 
\label{Perform thematic analysis of the challenges and critical
success factors} To identify the challenges and recommendations documented in the literature, we utilized thematic analysis - a standard data analysis method for qualitative data - to process the raw textual data ( challenges and recommendations). In particular, we adhered to the process outlined by Braun and Clarke \cite{braun2006using} by using  Nvivo - a tool designed for qualitative data analysis   \cite{braun2006using}. After extracting the textual data into an Excel spreadsheet, we imported the data into Nvivo to perform open coding, which involves breaking down the data into smaller components and assigning labels to each component \cite{sbaraini2011grounded}. This process was conducted iteratively, with codes generated in the initial stage being modified and updated in later stages.
Examples of the data analysis process for challenge and guideline identification are shown in Fig. \ref{Identification of the challenges using thematic analysis} and Fig. \ref{Developement of guidelines from thematic analysis} respectively.

From the analyzed data, we identified 20 challenges pertaining to the design, implementation, and evaluation phases of phishing education, training, and awareness interventions. We classify the challenges into three broad categories: first, design challenges relate to the concept, functionality, feature, or user interface of anti-phishing interventions. Examples of design challenges include \textit{inconsistent UI design across browsers and mobile devices} [P22, P49], \textit{unsuitable warning placement} [P2, P3, P5, P7, P11, P15, P25], and \textit{complex interface and configuration in the game design} [P28]. Second, implementation challenges relate to intervention automation, deployment, and adoption (e.g., \textit{the interdependancy between different factors and platforms in implementing effective anti-phishing measures} [P23, P38, P50]). Third, evaluation challenges relate to the measurement of usability and effectiveness of anti-phishing interventions and quantification of training outcomes. As developers and practitioners are the primarily targeted user groups of our personalized guidelines, we excluded challenges relevant exclusively to researchers (i.e., \say{Ch17 - limited demographic consideration in the study settings} [P1, P13, P14]). 

We collected several suggestions or recommendations documented in the literature such as \say{feedback should be provided whenever possible, regardless of the actual system reliability level [P33]}, \say{designing an effective security warning by providing better risk perception is a significant part [P20]}. We performed the thematic analysis of these recommendations to report 41 guidelines.

\begin{figure}[t] 
\centering
\includegraphics[trim=15 338 224 0,clip, scale = 0.34]{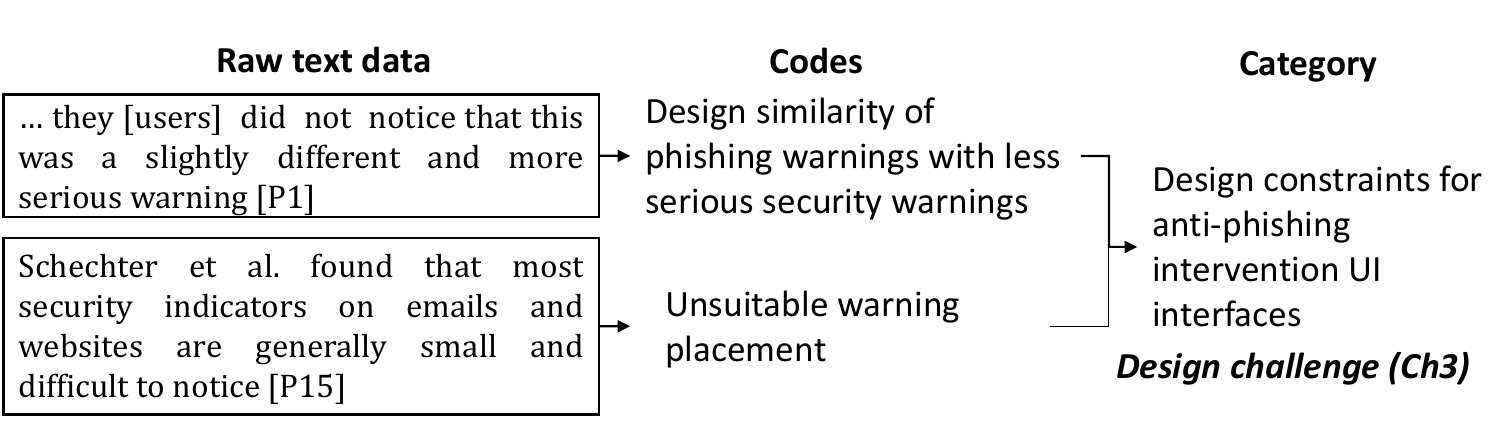}
 \caption{Identification of the challenges using thematic analysis}
 \label{Identification of the challenges using thematic analysis}
 \end{figure}
\begin{figure}[t] 
\centering
\includegraphics[trim=11 245 80 5,clip, scale = 0.30]{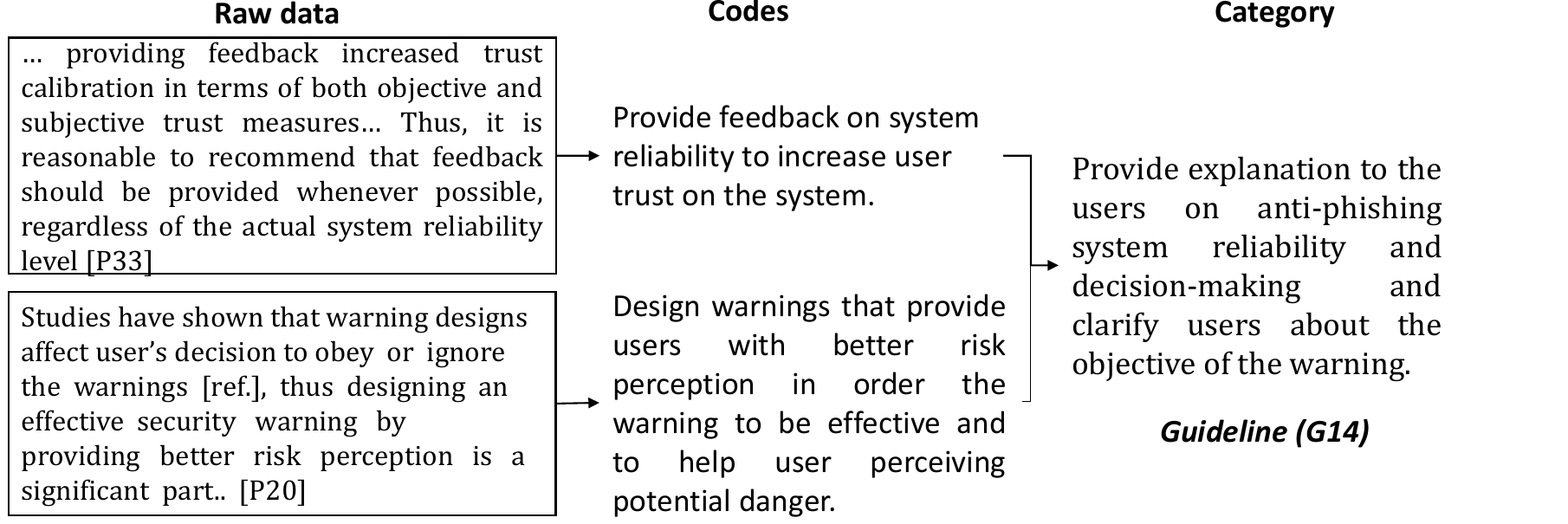}
 \caption{Identification of guidelines using thematic analysis}
 \label{Developement of guidelines from thematic analysis}
 \end{figure}

\subsection{Identifying practitioner groups}

\begin{figure}[t] 
\centering
\includegraphics[trim=0 140 170 20,clip, scale = 0.30]{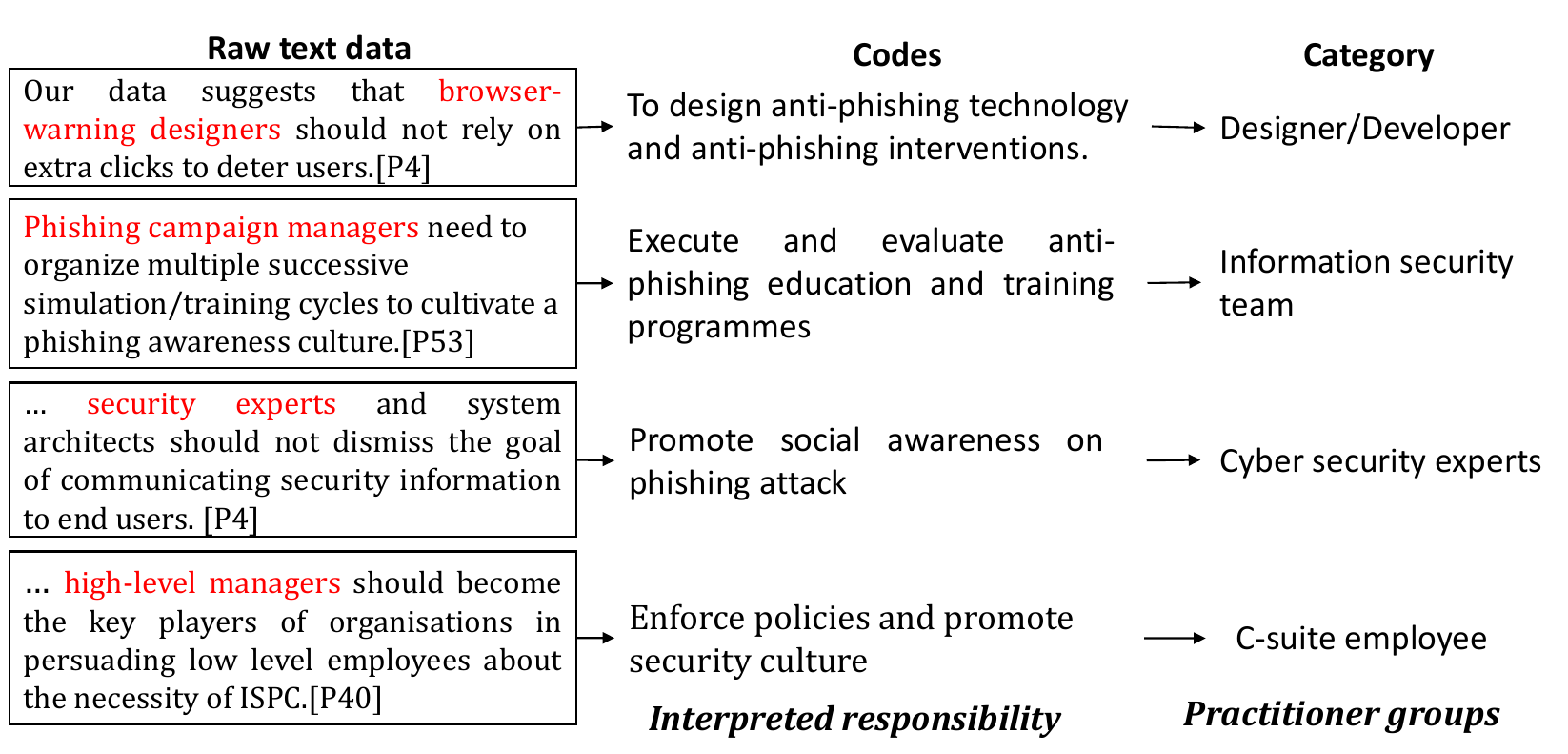}
 \caption{Categorization of different practitioner groups using thematic analysis}
 \label{Categorization of different practitioner group}
 \end{figure}


To report personalized guidelines for anti-phishing intervention developers and  practitioners, it is important to identify their specific needs and responsibilities involved in the design, implementation, and evaluation processes of anti-phishing interventions. 
To derive a mapping between practitioner groups and their roles and responsibilities, following existing studies in other domains (e.g., \cite{bhatt2020explainable,hong2020human,tomsett2018interpretable}) we categorize different practitioner groups based on their functional roles as documented in the literature.
From the literature we first gathered information on various practitioner roles, specifically 28 different groups, such as browser developer [P3, P8], platform designer [P4], designers of anti-phishing applications [P18], game developer [P36], information security officer [P27], chief information security manager [P34], and cyber security practitioners and decision makers [P53].
We then carefully investigated the responsibilities performed by each practitioner groups based on the recommendations collected from the MLR. For example, from the following textual data  \say{\textit{phishing campaign managers need to organize multiple successive simulation/training cycles to cultivate a phishing awareness culture [P53]}}, we infer that one of the responsibilities of \textit{phishing campaign managers} is to execute and evaluate anti-phishing education and training programs. Fig. \ref{Categorization of different practitioner group} shows an example of how we identified the responsibilities of different practitioner groups from raw textual data in the literature.

Based on these identified responsibilities of each practitioner groups, we then categorized all practitioners involved in anti-phishing interventions into four major categories, namely \textit{designer/developer}, \textit{information security team}, \textit{cyber security experts}, and \textit{C-suite employees}. Table \ref{Identified practitioner group and their resonsibilties} summarizes the responsibilities of different practitioner groups. A fraction of the responsibilities  overlap across different practitioner groups. For example, a designer/developer whose main responsibility is to \textit{design} anti-phishing tools may also  \textit{evaluates} an anti-phishing software [P22, P57, P61, P66, P67], because a designer/developer may need to test the usability of interventions before finalizing the design. However, as shown in Table \ref{Identified practitioner group and their resonsibilties}, evaluating anti-phishing interventions is mainly carried out by organizations' information security teams.

\subsection{Identifying and classifying interventions}
For each study collected in our MLR, we carefully examined the introduction, methodology, and results sections in search of any interventions  that were suggested, debated, or evaluated.   
Consequently, in order to tailor the guidelines to each type of intervention, we only included those treatments that had pertinent recommendations reported in the research. We classified the interventions into three types - education, training, or awareness - based on characteristics such as their intended goal, presentation, and method of delivery of anti-phishing information to users (the terms phishing education, training and awareness are defined in our online appendix \cite{supplementary}). We identified 2 types of phishing education: \textit{anti-phishing instructions} and \textit{educational games}. Identified phishing training interventions are: \textit{ phishing simulation and embedded training}, \textit{phishing training game}, \textit{narrative-based training}, \textit{instructor-based training}, \textit{information and guidance-based training}. For phishing awareness interventions, we found \textit{email client phishing indicators}, \textit{browser SSL warnings}, \textit{browser EV certificate warning}, \textit{browser security toolbar}, \textit{browser phishing warning}, \textit{QR code scanner phishing warnings} and \textit{Interactive custom phishing indicator}. Please refer to  our online appendix \cite{supplementary} for the definitions of each intervention.

\subsection{Identifying dominant factors} \label{Identifcation of dominant factors}
From the raw texts collected on the challenges and recommendations relating to anti-phishing interventions (as discussed in Section \ref{Perform thematic analysis of the challenges and critical
success factors}), we investigated the main \textit{reasons} for poor outcomes of existing anti-phishing interventions, as well as areas of improvement suggested by the authors to achieve better user experience. Based on this synthesized information, we identify and coin the term \textit{dominant factors}, which refers to  the individual, technical or organisational factors that may either enhance or impede the overall outcome of anti-phishing interventions. These factors are called \textit{dominant} as these factors were argued to influence the outcome of the anti-phishing interventions after empirical evaluation with the users in the results and discussions of our collected studies.  We adopt the terminology utilized by prior researchers \cite{dupont1997dirty, desolda2021human, hidellaarachchi2022influence} to designate the dominant factors identified in our investigation.

As an example of identifying a dominant factor, from the textual data  \textit{\say{staff may expect to learn more from experts while [college] students may expect to learn more from their peers [P48]}}, we derived an understanding that, according to the authors, the education qualifications of users have a significant bearing in determining their preference for the type of training methods and their effectiveness. This information indicates that designers could consider the educational qualifications of the end-users to improve user experience with the phishing intervention. Accordingly, the dominant factor identified here is users' \textit{educational qualification}. Fig. \ref{Dominant factor identification} illustrates an example of dominant factor identification from the raw textual data.

Fig.~\ref{meta model} depicts an example of the interconnection among challenges, guidelines, practitioner groups, interventions, and dominant factors derived from the raw text data. To determine the interconnection between challenges and interventions, we identify the interventions discussed in the study and the limitations mentioned within those interventions. Similarly, in establishing the interconnection among guidelines, interventions, and practitioner groups, we search for recommendations that are directed towards practitioners to enhance the outcomes of specific interventions. 

\begin{table*}[t!]
\centering
 \caption{Identified practitioner groups and their responsibilities}
\label{Identified practitioner group and their resonsibilties}
\footnotesize
\resizebox{\textwidth}{!}{%
\begin{tabular}{p{0.01\textwidth} p{0.10\textwidth}p{0.92\textwidth}}
\\\toprule
\textbf{No} & \textbf{Practitioners} & \textbf{Responsibilites/Activities} \\
\midrule
U1 & Designer/\newline Developer & \textbullet\ Design and deploy anti-phishing technology and anti-phishing interventions [P4, P19, P54, P60, P61, P65].
\newline \textbullet\ Conduct usability testing of anti-phishing intervention to improve the design [P22, P57, P61, P66, P67].
\\ \hline

U2 & Information \newline security team & \textbullet\ Implement the anti-phishing technology and anti-phishing solutions within the organization [P26, P50, P53 P61, P63, P67].
\newline \textbullet\ Execute and evaluate anti-phishing education and training programmes [P30, P54, P59, P61, P62, P69].
\\ \hline

U3 & Cyber security \newline experts&  \textbullet\ Make decisions on the appropriate elements and aspects to be included in the anti-phishing interventions [P13, P41, P42], promote social awareness on phishing attack [P4, P21].
\\ \hline

U4 & C-suite \newline employee & \textbullet\ Enforce policies to educate and train employees against phishing attack [P11, P38, P40 P50, P53, P54, P57, P67] and to promote security culture [P21, P26, P38, P40, P50,  P56, P57, P59, P60, P61, P67, P68, P69] within the organization.
\newline \textbullet\ Collaborate with the organization’s security team to adopt strong anti-phishing measures and prepare and execute a phishing incident response plan [P53, P56, P57, P60, P61].
\\\hline

\end{tabular}
}
\normalsize
\end{table*}%

\normalsize

 \begin{figure}[t] 
\centering
\includegraphics[trim=5 274 260 10,clip, scale = 0.41]{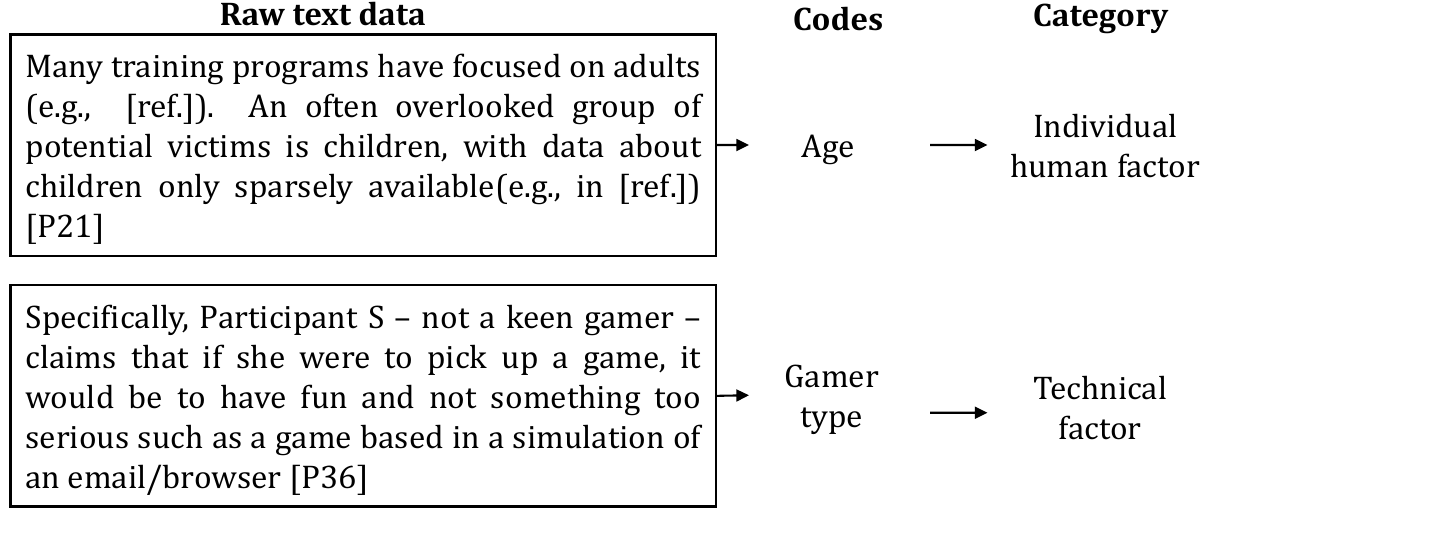}
 \caption{Dominant factor identification from raw text data}
 \label{Dominant factor identification}
 \end{figure}

 \begin{figure}[t] 
\centering
\includegraphics[trim=41 125 302 7,clip, scale = 0.38]{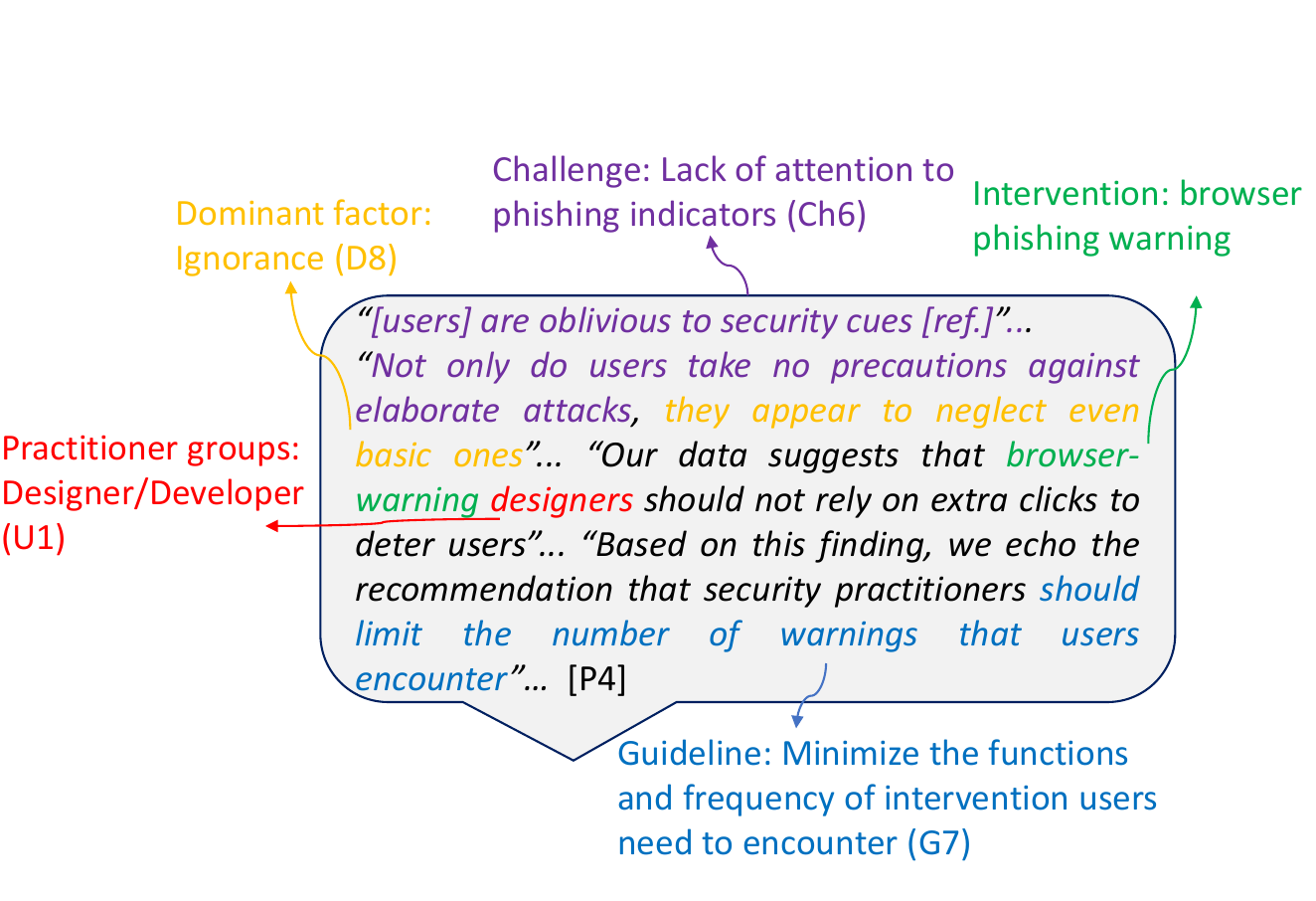}
 \caption{Example interconnection among challenges, guidelines, practitioner groups, interventions and dominant factors}
 \label{meta model}
 \end{figure}

\section{Research Findings}

\subsection{Dominant factors impacting anti-phishing interventions}
We analyzed the human factors discussed by Dupont \cite{dupont1997dirty} and refined by Desolda et al. \cite{desolda2021human} in the context of phishing attacks. A total of 8 human-oriented factors identified from our textual data (complacency, distraction, lack of communication, pressure, lack of knowledge, lack of resources, fatigue, and norms) confirm the factors documented by Dupont. A number of additional factors have emerged from our textual data, prompting us to search the literature for examples of their manifestation in order to categorize these factors. Following the approach by a study in the field of software engineering by Dulaji et al. \cite{hidellaarachchi2022influence}, we were able to identify and name the other previously unexplored factors, which make a significant contribution of new knowledge to this field. 

We identified 22 dominant factors to underscore the significance of incorporating user needs and preferences in the design, implementation, and evaluation of anti-phishing interventions. These dominant factors emphasize that neglecting user's requirements and inclinations can hinder practitioners' efforts in providing tailored design, implementation, and assessment of a system, which in turn results in compromised usability and suboptimal outcomes. Table \ref{table: Dominant factors and their impact on PETA} details our 22 identified dominant factors and how their inclusion or absence can influence the outcome of anti-phishing interventions. We grouped our dominant factors into three categories: individual human factors, technical factors, and organizational factors.

\subsubsection{Individual human factors}
Based on the analysis of selected studies, we identified that various demographic characteristics of individual users require greater attention from developers and practitioners to enhance the efficacy of anti-phishing interventions across different user groups. According to the literature, developers and practitioners need to take into consideration a number of user demographic characteristics, including \textit{age} and \textit{educational qualification}, in order to enhance the efficacy of anti-phishing interventions across different user groups.

In addition to demographic characteristics, our additional dominant factors also relate to cognitive functioning and limitations of individual users in order to provide a comprehensive picture of user-level characteristics. Specifically, \textit{knowledge decay}, \textit{distraction}, \textit{lack of attention} and \textit{lack of motivation} all constitute individual limitations that can reduce user's capacity to effectively identify or counteract phishing attacks. For example, studies have shown that the knowledge acquired by individuals through phishing training tends to degrade or dissipate over time [P7, P13, P21, P31, P34]. Again, phishing warnings frequently pass unnoticed as users become diverted by concurrent tasks or are incapable of maintaining attention across multiple stimuli [P13, P14, P41].

Our investigation reveals that certain personality traits or user characteristics, such as \textit{complacency} and \textit{optimism bias}, may lead to users disregarding anti-phishing interventions. For example, users tend to overestimate the efficacy of their organization's anti-phishing measures [P7] and exhibit over-reliance on website content that is visually attractive [P2, P5, P8, P11, P49]. It is noteworthy to mention that our dominant factors represent distinct attitudes or mindsets. For example, complacency involves a belief that things are good enough as they are and that there is no need for further effort or change \cite{nwankpa2023remote} which may cause complacent individuals to overlook potential phishing risks. Conversely, optimistic individuals feel less likely to experience cyber attacks which may cause them to share passwords, visit untrustworthy websites and so on \cite{alnifie2023appraising}.

An absence of tailored design, implementation, and evaluation of anti-phishing interventions by taking into consideration user needs and preferences could result in overwhelming the user and causing \textit{security fatigue}. For example,  fatigue can result from frequent exposure to warning and risk notifications, thereby reducing their effectiveness [P4, P13, P14, P17, P18, P26]. Similarly, excessive training can lead to training fatigue [P34, P52, P53, P58, P60, P61, P62, P69]. The complexity inherent in software installation procedures, as well as the intricacy of the process for reporting phishing incidents, may lead to \textit{lack of user motivation}.

\subsubsection{Technical factors}
The effectiveness of anti-phishing warnings is greatly influenced by the \textit{type of devices} utilized by the users. 
 \onecolumn
\footnotesize
 \setlength{\tabcolsep}{5pt}
 \begin{longtable}{p{0.3cm} p{1.7cm}p{13.8cm}p{0.5cm}}
 \caption{Dominant factors and their impact on anti-phishing interventions}
\label{table: Dominant factors and their impact on PETA}\\
 \hline \multicolumn{1}{c}{\textbf{No}} & \multicolumn{1}{c}{\textbf{Factors}} & \multicolumn{1}{c}{\textbf{Impact}} & \multicolumn{1}{c}{\textbf{\#}}
  \\ \hline 
 \endfirsthead
 \multicolumn{4}{c}%
 {{ Dominant \ factors \ and \ their \ impact \ on \ PETA -- continued from previous page}} \\
 \hline \multicolumn{1}{c}{\textbf{No}} & \multicolumn{1}{c}{\textbf{Factors}} & \multicolumn{1}{c}{\textbf{Key points (included papers)}} & \multicolumn{1}{c}{\textbf{\#}}  \\ \hline 
 \endhead
 \hline \multicolumn{4}{c}{{Continued on next page}} \\ \hline
 \endfoot
 \hline
 \endlastfoot
 \hline
 \multicolumn{4}{c}{\textbf{Individual human factors}} \\
 \midrule
 D1& Age & \textbullet\  Children aged 8-13 require specialized phishing educational intervention as they are biologically less attentive [P21, P35]. 
 \newline \textbullet\ Teenagers tend to make decisions quickly without considering the consequences, and are more susceptible to being persuaded by urgency and panic-inducing phishing emails [P35]. \newline \textbullet\ Older employees have relatively bad training outcomes as they prioritize maintenance over growth [P40]. \newline \textbullet\ Age 18-25 are vulnerable to phishing attacks [P6].  & 4  \\\cline{1-4}

 D2 & Complacency & \textbullet\ Users' overconfidence in the appealing web content leads them to disregard phishing warnings [P2, P5, P8, P11, P49]. \newline \textbullet\ Users' prior experience with websites results in overconfidence, causing them to disregard phishing warnings [P1, P2, P5, P11, P13, P25, P44]. \newline \textbullet\ Users over-rely on site reputation and trust the warning [P14].  \newline \textbullet\ Users are overconfident about their ability to detect phishing [P43, P44, P45], over-trust on their organizational technical phishing solutions [P7]. & 14  \\\cline{1-4}

 D3 & Confusion & \textbullet\ User confusion arises due to similarity in domain names [P45], webhosting [P45, P49], distinct warning design patterns among vendors [P25, P49] and conflicting information present in the anti-phishing guidelines [P42]. \newline \textbullet\ Users become confused about the purpose of a received training email [P5]. & 5  \\\cline{1-4}

D4 & Curiosity &  \textbullet\ Users click on the phishing link out of curiosity [P2, P25].  & 2  \\\cline{1-4}

D5 & Distraction &  \textbullet\ Users are distracted by other tasks as security is not their main concern [P13, P14, P41]. \newline \textbullet\ Individuals are unable to focus on multiple things simultaneously (e.g., noticing on phishing warning while doing online shopping) [P13]. & 3  \\\cline{1-4}

D6 & Educational Qualification &  \textbullet\ Phishing stories from a peer is an effective method of training for college students [P48].\newline \textbullet\ University staffs learn better from facts from an expert-based training method [P48]. \newline \textbullet\ Compared to bachelor’s degree, users having master’s and PhD degrees are more confident in detecting phishing [P52]. & 2  \\\cline{1-4}

D7 & Knowledge decay &  \textbullet\ The knowledge gained by users during phishing training tends to dissipate over time [P7, P13, P21, P31, P34]. & 5  \\\cline{1-4}

D8 & Ignorance &  \textbullet\ Users failed to look at anti-phishing interventions [P7, P13, P17, P28], ignored as web content looked legitimate [P2] and when received a high frequency of warnings [P4].  & 6  \\\cline{1-4}

D9 & Lack of communication &  \textbullet\ Before designing and implementing anti-phishing software, users' interests and needs are not well investigated [P16]. & 1  \\\cline{1-4}

D10 & Lack of motivation &  \textbullet\ Users are not motivated enough to install anti-phishing software on their devices [P31, P37], show unwillingness to report phishing due to a complicated reporting process [P50, P58, P63] and do not find the training and educational material interesting [P10, P19, P28]. & 8  \\\cline{1-4}

D11 & Lack of trust &  \textbullet\ Users do not trust anti-phishing warnings due to limited accuracy of anti-phishing tools [P1, P11]. & 2  \\\cline{1-4}

D12 & Optimism bias &  \textbullet\ Optimistic users tend to be less conscious as they believe that negative events only happen to others [P13].& 1  \\\cline{1-4}

D13 & Perceived vulnerability and severity &  \textbullet\ An individual's heightened understanding of the consequences of phishing attacks enhances their resistance to these types of attacks [P40].& 1  \\\cline{1-4}

D14 & Pressure &  \textbullet\ Phishing incident response by IT staff gets delayed due to the reception of a high volume of phishing reports [P50]. \newline \textbullet\ An individual receiving a high volume of emails is more susceptible to phishing attacks [P26].& 2  \\\cline{1-4}

D15 & Fatigue &  \textbullet\ Providing comprehensive instruction could result in overwhelming the user [P13]. \newline \textbullet\ Frequent exposure to warning causes warning fatigue [P4, P13, P14, P17, P18, P26]. \newline \textbullet\ Frequent risk notifications and excessive training result in training fatigue [P34, P53, P58, P60, P61, P62, P69]. & 13  \\\cline{1-4}

\multicolumn{3}{c}{\textbf{Technical factors}} \\
 \midrule
 
D16 & Device type  &  \textbullet\ Individuals who rely on mobile devices are at a higher risk, as phishing signs are obscured or not fully visible on the small screens of mobile devices [P49].  & 1  \\\cline{1-4}

D17 & Gamer type &  \textbullet\ A casual player is unsatisfied with playing a phishing game that is designed for serious gamers, and conversely, a serious gamer is unfulfilled playing a phishing game that is intended for casual players [P36].  & 1  \\\cline{1-4}

D18 & Lack of knowledge &  \textbullet\ Users do not understand anti-phishing warnings due to lack of knowledge about security and security indicators [P1, P4, P5, P6, P7, P8, P9, P10, P11, P13, P14, P17, P20, P21, P28, P35, P39, P46, P47, P49].& 20  \\\cline{1-4}

D19 & Lack of resource &  \textbullet\ User do not have enough infrastructure support when they work from home [P6]. \newline \textbullet\ Absence of abstractness in the anti-phishing recommendations and lack of advanced anti-phishing tools reduces users’ self-efficacy [P42]. \newline \textbullet\ Users do not receive training emails due to emails being in the spam folder [P28]. & 3  \\\cline{1-4}

\multicolumn{3}{c}{\textbf{Organizational  factors}} \\
 \midrule
D20 & Organizational position &  \textbullet\ Employees in a higher position in an organization are more vulnerable regardless of the phishing training or punishment [P40].  & 1 \\ \cline{1-4}

D21 & Social influence &  \textbullet\ People trust others' phishing stories as they perceive this information as trustworthy [P15]. \newline \textbullet\ Observing others share information results in heightened levels of disclosure [P13]. \newline \textbullet\ The motivation, self-efficacy, and cognitive ability of employees are impacted by the social relationships within and surrounding the organization [P26, P40]. & 4 \\ \cline{1-4}
D22 & Norms&  \textbullet\ Organization’s procedural measures (e.g., security policies, standards and guidelines) have a beneficial effect on raising security consciousness [P38].  & 1 \\ \cline{1-4}

 
 \end{longtable}
\begin{multicols}{2}
 \normalsize 
Research has shown that web developers tend to avoid adding phishing indicators to mobile browsers to save space for web content [P16]. Based on a research report by [P36], it is evident that anti-phishing educational games fail to tailor their content to the specific interests of individual user groups. To elaborate, the games designed for serious gamers do not meet the expectations of casual players, and vice versa. Consequently, designers of anti-phishing games ought to take into account the \textit{type of gamers} as a factor when creating an educational game that can cater to the unique requirements of both casual and serious gamers.

According to multiple studies, the challenge faced by users with limited technical knowledge is attributable to their insufficient familiarity with security indicators and tools, as well as the complexity of the requirements of third-party tools [P1, P4, P5, P6, P7, P8, P9, P10, P11, P13, P14, P17, P20, P21, P28, P35, P39, P46, P47, P49]. 
This highlights the significance of incorporating the needs of novice users into the design process. Additionally, the effectiveness of users in detecting and preventing phishing attacks is reduced by a lack of resources, such as infrastructure support and advanced anti-phishing tools, and the absence of abstractness in anti-phishing recommendations [P6, P42].

\subsubsection{Organizational factors}
We identified several organizational factors from the literature that could be incorporated to enhance the effectiveness of the anti-phishing interventions. For example, according to research, \textit{higher-positioned} employees in an organization are more susceptible to phishing attacks, regardless of their previous training or negative experience of being victims of phishing attacks  [P40].  

The literature shows that the implementation of security policies, standards, and guidelines in an organization is beneficial to increasing phishing awareness among employees. Additionally, the positive impact of organizational \textit{norms} and \textit{normative beliefs} on employees is observed in their exercise of greater caution, when opening potentially harmful phishing emails [P38].  Employees' motivation, self-efficacy, and cognitive ability are affected by social relationships [P26, P40]. For example, according to several studies [P13, P15], individuals tend to trust and share phishing stories based on the perceived trustworthiness of the information and the influence of social relationships within and around their organization.

\subsection{Guidelines for the design, implementation and evaluation of anti-phishing interventions}
\label{Guidelines for the design, implementation and evaluation of PETA}
We present our guidelines and the rationale underpinning each guideline in Table \ref{table:Guidelines for PETA}. In the following paragraphs, we briefly discuss the guidelines personalized to different practitioner groups, intervention stages, intervention types, challenges and dominant factors in anti-phishing interventions.

\textbullet\ Among the 41 guidelines, 20 are mapped as relevant to our first user group (U1) consisting of designers and developers. This user group has the highest number of guideline relevant to them. These guidelines (G1 to G11, G13, G14, G16, G17, G19, G21 to G23, G27) include recommendations on interface design, placement of the phishing indicator, intervention content design, user engagement strategies, attention-drawing techniques, and enhancements for existing and future intervention designs.

\textbullet\ Our second user group (U2) consisting of \textit{information security teams} of the organizations have a total of 19 guidelines (G12, G15, G16, G18, G20, G24, G25, G28, G29, G31 to G38, G40, G41) mapped to them as potentially applicable. These guidelines are intended to assist these cyber security professionals in conducting effective phishing education and training sessions, implementing measures to reinforce the organization's security, enhancing the speed and efficiency of phishing incident response and reporting procedures, as well as improving the educational resources made available to users to better cope with the threat of phishing.

\textbullet\ The guidelines G25 and G26 have been devised for \textit{cyber security experts} (U3)), with a primary focus on conducting an investigation on the anti-phishing solution before adoption, as well as implementing protocols to ensure organizational adherence to a standardised email and anti-phishing webpage template. 

\textbullet\ The guidelines G26, G27, G30, and G39 have been reported specifically for \textit{C-suite employees} (U4) within organizations. While these guidelines are primarily intended for use by C-suite employees, certain guidelines, such as G28 and G29, are also relevant to the organization's security team (U2). These guidelines emphasize the importance of collaboration between the C-suite and the security team to develop policies that meet the needs of employees and provide better support for victims of phishing attacks. 

\textbullet\ The user group U1 is mapped to guidelines for all three stages of phishing intervention, whereas the user group U2 are linked to guidelines solely for the implementation and evaluation stages. In contrast, user groups U3 (cyber security expert) and U4 (C-suite employee) have guidelines exclusively for the implementation stage of phishing intervention.

\textbullet\ Across the three stages of anti-phishing interventions, we provide 19 guidelines (G1 to G11, G13, G14, G16, G17, G19, G21, G22, and G27) for the design stage, 17 guidelines (G12, G15, G16, G18, G20, G23 to G28, G30 to G32, G35 to G37, and G39) for the implementation, and 8 guidelines (G13, G16, G29 to G33) for the evaluation of anti-phishing interventions. It is noteworthy that some guidelines are applicable to multiple different stages. For example, G16 is applicable to the design, implementation, and evaluation stages. We arrived at this recommendation based on several studies (e.g., P7, P13, P15, and P16) that suggest incorporating users' preferences in the design (e.g., the layout of anti-phishing intervention), implementation (e.g., training methods used to educate users), and evaluation (e.g., email templates used to assess users' phishing knowledge) of phishing interventions.

\textbullet\ The guidelines presented in this study have the potential to address several challenges in the design, implementation, and evaluation of existing anti-phishing interventions. For example, these guidelines can be particularly useful in addressing design constraints specific to anti-phishing warning user interface (e.g., G7, G8, G9), improving content-related issues for phishing education and training (e.g., G4, G5, G6), mitigating issues related to the deployment and adoption of anti-phishing technologies (e.g., G18 and G7), overcoming limitations associated with existing anti-phishing planning, policies, and guidelines (e.g., G12, G24, G25). 

\end{multicols}
\onecolumn
\footnotesize
 \setlength{\tabcolsep}{5pt}
 \begin{longtable}{p{0.2cm}p{6cm}p{10.6cm}}
 \caption{Guidelines for anti-phishing interventions }
\label{table:Guidelines for PETA}\\
 \hline \multicolumn{1}{c}{\textbf{No}} & \multicolumn{1}{c}{\textbf{Guidelines}} & \multicolumn{1}{c}{\textbf{Rationale}}
 \endfirsthead
 \multicolumn{3}{c}%
 {{ Guidelines \ for \ anti-phishing interventions -- continued from previous page}} \\
 \hline \multicolumn{1}{c}{\textbf{No}} & \multicolumn{1}{c}{\textbf{Guidelines}} & \multicolumn{1}{c}{\textbf{Rationale}}  \\ \hline 
 \endhead
 \hline \multicolumn{3}{c}{{Continued on next page}} \\ 
 \endfoot

 \endlastfoot
 \midrule
G1 & Remove deceptive user interface elements for unverified emails and incorporate an alert icon within the email client to indicate potentially fraudulent emails.  & \textbullet\ Disabling misleading UI elements (e.g., profile photo, email history) for unverified sender addresses will reduce user confusion [P16]. \newline \textbullet\ Placing a security indicator for unverified email delivered to the user acts as a forcing function for the sender domain to configure their SPF/DMRC/DKIM correctly [P7, P16].  \\\cline{1-3}

G2 &  Clearly display the underlying URL of a suspicious link in the email client & \textbullet\ Clearly displaying the underlying URL of a suspicious link in the email client (link-focused warning) make it easier for users to notice where the links' actual destination [P25]. \\ \cline{1-3}

G3 & Incorporate progressive disclosure in the design and add a learn more button. & \textbullet\ Progressive design and learn more buttons help to facilitate general advice, satisfy user curiosity, and support user investigations [P4, P5, P25, P51]. \\\cline{1-3}
 
G4 & Use visual examples and explanations and avoid technical jargon in the content. & \textbullet\ Avoiding technical details in the content can make them understandable to non-expert users [P1]. \newline \textbullet\ Integrating visual examples and explanations on phishing cues presented helps users memorize and understand better [P42]. \\\cline{1-3}

G5 & Present abstract information and leverage situated learning in the content. & \textbullet\ Leveraging situated learning in the design can make the intervention interesting and engaging, and also improves learning outcomes [P5, P10, P19, P28, P34, P36, P37, P61, P62]. \newline \textbullet\ Too much information in the content can be unappealing to inexperienced users [P1, P5, P13, P18, P41]. \newline \textbullet\ Adopting situated learning is beneficial as learning science suggest that simply asking users to follow some advice would not be helpful [P5].\\\cline{1-3}

G6 & Introduce varieties in the content and keep the information up to date. & \textbullet\ Including varieties in the content can help users tackle new and emerging phishing attacks [P19, P57, P58, P59, P61, P65]. \\\cline{1-3}

G7 & Minimize the functions and frequency of intervention users need to encounter. & \textbullet\ Limiting the frequency of the warnings reduce warning fatigue [P4]. \newline \textbullet\ Minimum number of functionalities in the game can help finish the game easily, easy for users to remember when functionalities are less [P10]. \\\cline{1-3}

G8 & Design phishing warnings differently from standard warnings.  & \textbullet\ Variation in the design increases the likelihood for users to read it, ensures they are taken seriously and prevent habituation [P1, P2, P14]. \\\cline{1-3}

G9 & Make the critical information easily accessible and visible to the users. & \textbullet\ To make users easily notice the warnings [P1, P4, P8, P25], increase warning adherence [P25] and to impose forced attention [P8, P25]. \\\cline{1-3}

G10 & Create uniform phishing indicators across different browsers and mobile interfaces. & \textbullet\ This will reduce the susceptibility of mobile device users [P16]. \\\cline{1-3}

G11 & Provide users clear choices and actionable items to proceed. & \textbullet\ Active interruption and actionable items minimize the user’s workload, are naturally noticeable and users can use their time efficiently [P1, P2, P4, P5, P7, P20, P22, P24, P25 P41, P43, P44] \\\cline{1-3}

G12 & Offer intervention immediately after users fall for phishing. & \textbullet\ Avoiding delay in displaying warnings minimizes users’ confusion [P5]. The right timing of training intervention provides instant education [P2]. \\\cline{1-3}

G13 & Perform usability tests and collect user feedback. & \textbullet\ Collecting users’ feedback from usability testing can improve future intervention design [P18, P22, P57, P61, P66, P67].  \\\cline{1-3}

G14 & Provide an explanation to the users on anti-phishing system reliability and decision-making and clarify users about the objective of the intervention. & \textbullet\ Feedback on the anti-phishing system increases users' trust [P7, P8, P11, P14, P33, P39, P43], helps users perceive potential danger [P20], increases user understanding and improves user ability to detect phishing [P18, P39]. \newline \textbullet\ Making it clear to the users why they have displayed the intervention or not taken to the website to avoid their confusion [P5,P14]. \\ \cline{1-3}

G15 & Use both technical and human-centric defence mechanisms to cope with phishing. & \textbullet\ Prevent user’s over-reliance on technology, provide additional defence in detecting unpredictable, highly dynamic, and increasingly sophisticated phishing attacks
[P3, P5, P12, P17, P18, P26, P27, P28, P38, P41, P51, P53, P57, P58, P59]. \newline \textbullet\ Educating users about the security properties of different interventions remove their misunderstanding that leads to mistake [P14].\newline \textbullet\ Training all individual who has access to the organization increase the organization’s robustness [P53]. \newline \textbullet\ Human-centric defence mechanisms organized by C-suit employees can help low-level employees in the organization to learn about phishing [P21, P38, P40, P56, P57, P59, P61, P67, P68, P69]. \\\cline{1-3}

G16 & Personalize the intervention style and medium based on the target user’s demographic. & \textbullet\ Personalized phishing training can take into account user’s preferences (e.g., individual preferred training method [P15, P21], content relevant to the organization [P16, P58], roles and responsibilities [P40, P53, P58, P60], age [P21, P35]) to ensure users receive targeted education and training [P7, P13, P15, P16, P21, P26, P35, P36, P40, P48, P52, P53, P57, P58, P59, P60, P61, P62, P64, P66, P67]. \\\cline{1-3}

G17 & Consider the decision-making process and vulnerabilities of humans in the design. &  \textbullet\ Taking into account the vulnerabilities and decision-making processes of the user (e.g., users’ misconceptions and perspectives [P11], perceived threat [P9]) increases the effectiveness of anti-phishing interventions for end users and assist to develop the tailored approach [P4, P6, P7, P9, P11, P18, P24]. \\\cline{1-3}
 
G18 & Configure IT system for phishing training. & \textbullet\ Preparing IT system to avoid simulated email being filtered by technical filters helps users being missed for training [P69]. \newline \textbullet\ Verifying  if inventory management software is utilizing scanning, analysis, or probing techniques help detect abnormally high levels of external IP addresses [P54]. \\\cline{1-3}

G19 & Design visually distinct user-friendly URL bar. & \textbullet\ Noticeable and consistent URL bar helps users differentiate legitimate and malicious domains easily [P2, P8, P46]. \\\cline{1-3}

G20 & Use automated platforms and improved tools for phishing training, incident management and reporting. & \textbullet\ Automated approaches help to better support managing complex situations, delivering personalized content and threat identification [P61, P63, P67, P50]. \\\cline{1-3}

G21 & Disable JavaScript on login forms when a form element is in focus. & \textbullet\ Deactivating JavaScript on webpages every time the focus is put on a form element prevents the attacker from capturing the keystrokes or initiating timing attacks [P16, P22, P23]. \\\cline{1-3}

G22 & Explain the capabilities and effectiveness of the deployed anti-phishing solution clearly to the users. & \textbullet\ Reliable trust signals to the users can prevent over-trust and over-reliance on the deployed anti-phishing solutions [P11].
\newline \textbullet\ Utilizing interactive error messages to elucidate the purpose of a website can deter users from engaging in destructive actions [P43, P44]. \\\cline{1-3}

G23 & Use email authentication protocols to encrypt emails and filter out incoming malicious emails. & \textbullet\ To achieve better resiliency [P18,P51] and to make more informed decision [P16, P27] on the incoming emails. \\\cline{1-3}

G24 & Send pre-notification to the users before conducting phishing training, however, perform random phishing training. & \textbullet\ Sending pre-notification to the participants prevents discomfort [P30, P69]. \newline \textbullet\ Emphasising on the anonymity of phishing training can reduce the effect of prairie dogging and estimate of organization’s likelihood to fall victim to phishing [P59, P61, P62, P69]. \\\cline{1-3}

G25 & Conduct prior investigation before adopting anti-phishing tools, identify most vulnerable group and determine priority topics. & \textbullet\ Perform prior research and analyze the reviews on tool vendors to select the right tool [P26, P61] \newline \textbullet\ Identifying vulnerable users can help reduce training time and efforts [P26]. \newline \textbullet\ Teaching everything or huge amount of information can cause security fatigue [P13]. \\\cline{1-3}

G26 & Follow a consistent template for organizational emails and create a standard template for anti-phishing webpages. & \textbullet\ A consistent email structure helps employees to notice the discrepancies in phishing emails easily [P41]. \newline \textbullet\ A standardized template for anti-phishing webpages reduces inconsistency helps avoid confusion  and helps web-designer implement their anti-phishing tools easily [P42]. \\\cline{1-3}

G27 & Introduce a user-friendly, built-in phishing reporting tool within the client system. Develop a formal procedure to handle phishing reports. & \textbullet\ Having a formal procedure placed makes it convenient to handle phishing reports [P50]. \newline \textbullet\ An in-client phishing incident reporting tool makes phishing reporting easier [P58, P63]. \\\cline{1-3}

G28 & Get employees' feedback to modify the organization’s policy. & \textbullet\ Obtain staff's feedback after phishing simulation to modify the organization policy accordingly to meet staff's needs [P50]. \\\cline{1-3} 

G29 & Deploy help-desk and victim  support for users. & \textbullet\ Deploying post simulation help desk support allows further users’ investigations [P51]. \newline \textbullet\ Deploying help-desk support can assist external users in determining the authenticity of an email sent from the organization [P51]. \newline \textbullet\ Add a victim support option in the anti-phishing webpages can help users to fix potential problems [P42]. \\\cline{1-3}

G30 & Create a structured policy and documentation. Regularly assess and manage phishing awareness efforts. & \textbullet\ Appropriate policy and documentation ensure that all the employees adapt themselves to security countermeasures and requirements [P26, P38, P60]. \newline \textbullet\ Continuous measurement, improved management and policy making helps to achieve improved phishing defence [P11, P38, P40, P50, P53, P54, P57, P67]. \\\cline{1-3}  

G31 & Conduct phishing simulation with embedded training. & \textbullet\ Assist the organization's security team in practicing the handling and response to simulated phishing incidents to enhance preparedness for real phishing attacks [P53, P56, P57, P60, P61]. \newline \textbullet\ Embedding learning content with phishing simulation provides education on demand [P5, P7, P12, P27, P53, P56, P57, P58, P59, P60, P61, P67, P68, P69]. \\\cline{1-3}

G32 & Conduct phishing simulation that adheres to the guidelines of the data privacy policy appropriate to the region. & \textbullet\ Data privacy policy-compliant phishing training protects participants sensitive information, hence reducing data breaches [P26, P69]. \\\cline{1-3}

G33 & Provide users immediate feedback on their performance. & \textbullet\ Users feel motivated if instant corrective feedback is provided after testing and evaluating their phishing knowledge in their regular environment [P7, P10, P31]. \\\cline{1-3} 

G34 & Use realistic and equally difficult training emails. Use challenging questions to test phishing knowledge. & \textbullet\ Realistic and equally difficult email helps to test the persistence of the training’s effect [P7]. \newline \textbullet\ An extensive test with challenging questions reduce repetitive training costs and can help avoid the ceiling effect [P21]. \\\cline{1-3}

G35 & Implement progressive and self-adaptive phishing training. & \textbullet\ Dynamic and self-adaptive phishing training improve user sensitivity to deception cues [P24, P63, P64, P66]. \\\cline{1-3}

G36 & Adopt video and interactive education and training materials. &  \textbullet\ Video and interactive training are more effective as users do not need refreshment very quickly [P5, P11, P19, P34, P36] \\\cline{1-3}

G37 & Utilize the expertise of external service providers to aid in phishing knowledge assessment and awareness material development. & \textbullet\ Leveraging external service providers can support better phishing knowledge assessment and awareness material development [P54, P60]. \\\cline{1-3}

G38	& Choose evaluation metrics and baselines that are useful and relevant. & \textbullet\ Click-through rate should be normalized based on the persuasiveness of the training template to produce a sound analysis and evaluation [P32, P54, P56, P58, P59, P60, P61, P68]. \\\cline{1-3}

G39 & Train users how to report phishing and reward secure behaviour. & \textbullet\ Training users on how to report phishing incidents and explaining the benefits of reporting can help to establish a phishing reporting culture [P26, P50, P58, P60, P69]. \newline \textbullet\ Rewarding employees for their secure behaviour can motivate and encourage them to perform better [P30, P61, P66]. \\\cline{1-3}

G40 & Conduct multiple cycles of follow-up training. & \textbullet\ Help to assess users' short-term and long-term knowledge retention after training [P26, P31, P52, P54, P57, P58]. \newline \textbullet\ Repetitive training in a short period helps users learn a second time if they had difficulty understanding in the first time [P5, P7, P24, P27, P34, P53, P56, P57, P62, P67, P68, P69]. \newline \textbullet\ Follow-up training (for children) to counter knowledge decay of the ability to identify phishing [P21]. \\\cline{1-3}

G41 & Avoid frequent reminders and over-training and  keep the reminders short and simple. & \textbullet\ Avoiding frequent risk notifications and over-training reminders can reduce training fatigue [P34, P52, P53, P58, P60, P61, P62, P69]. \newline \textbullet\ Including a lower bound of information in the reminder measures can reduce security fatigue [P34].\\
\bottomrule
 
 \end{longtable}
%
\begin{multicols}{2}
\normalsize

\textbullet\ Among our 3 main intervention types (i.e., education, training and awareness), our analysis has yielded a set of 9 guidelines for education interventions (G5, G7, G16, G17, G25, G26, G29, G33, G36), 27 guidelines for training interventions (G5, G6, G7, G11 to G16, G18, G20, G24, G25, G27, G28, G30, G31 to G39), and 19 guidelines for awareness interventions (G1 to G5, G7 to G11, G13 to G17, G19, G21 to G23). 

\textbullet\ Some of our challenges (i.e., \say{Ch5 - 
 performance limitations of anti-phishing tools}, \say{Ch18 - insufficient usability and effectiveness evaluation of phishing interventions}, \say{Ch19 - lack of sophisticated quantification of phishing training outcome}) do not exhibit any discernible dominant factors. As a result, the guidelines G13 and G38, which are intended to address these challenges, are not linked to any particular dominant factor.

\section{Threat to validity}
The guidelines we compiled may not be \textit{comprehensive} because potential design, implementation and evaluation principles are not always explicitly articulated \cite{jampen2020don}. The presented guidelines have been formulated with specific regard to the collected study context being investigated, and therefore, their \textit{generalizability} may be limited. As a means of substantiating these standpoints, we posit that the 69 studies used in this research were collected through a rigorous process of quality assessment. Most of the guidelines that we have formulated are supported by more than one study in the literature, which underscores their applicability in contexts that are different from the collected study context. This is because these studies have involved diverse user types, varying sample sizes, different intervention types, and other variables. Additionally, our analysis encompassed industry reports [e.g., P58] and case studies with various organizations [e.g., P63, P64, P65]. The inclusion of grey studies facilitates the mitigation of bias that stems from a proclivity to publish studies that report favorable results exclusively \cite{kamei2021evidence}.

Regarding the \textit{representativeness} of the data used in this study, the collection of textual data is restricted to 53 academic and 16 grey studies which have been identified in literature searches. We recognize that future research could broaden the scope of the searches and analysis by including additional data sources such as interviews and surveys \cite{hermawati2016establishing}. In order to validate and strengthen the usability and effectiveness of our guidelines, we plan to conduct a semi-structured interview study with developers and cyber security practitioners in our future work.

As multiple researchers were involved in this study, in order to minimise \textit{researcher bias}, various activities (e.g., study selection, data search, extraction, analysis, and synthesis) were conducted in accordance with a well-defined research protocol, following the established guidelines proposed by Kitchenham and Charters \cite{kitchenham2007guidelines} and Garousi et al. \cite{garousi2019guidelines}. The research protocol was modified and updated by conducting a pilot study of randomly selected 10 studies. The first author collected 90\% of the data (62 out of 69), while the third author extracted the remaining 10\% (7 out of 69). All data were shared in a collaborative folder and cross-checked by each author and any issues or disagreements were resolved in weekly research meetings among the authors.

While employing thematic analysis permits the data analysis to be grounded in the textual data collected from academic and grey literature studies, there is a threat of \textit{subjectivity} of the data analysis \cite{rajapakse2021empirical}. To alleviate this threat, we discussed the issues and concerns of the emergent findings throughout the study in the weekly meetings. Throughout the iterative and intertwined rounds of data collection and analysis, the first author led the data analysis with support from other researchers who acted as validators at each stage. 
\section{Conclusion}
Current anti-phishing interventions encounter several obstacles, such as poor user interface design, lack of engaging and interesting content, incomplete or outdated anti-phishing instructions, flawed anti-phishing training implementation, and deficient anti-phishing policies within organizations. The usability issues that arise from the current one-size-fits-all anti-phishing interventions can be attributed to a need for greater awareness among developers and practitioners of end-user requirements and preferences. There is a current lack of available personalized guidelines to assist developers and practitioners for this purpose.

To address this gap, in this study, we first identified 22 dominant factors, consisting of 15 individual, 4 technical, and 3 organizational factors that impact the effectiveness and outcomes of anti-phishing interventions from 53 academic and 16 grey literature studies. We then present 41 guidelines to aid developers and practitioners in addressing these issues within current anti-phishing intervention design, implementation, and evaluation. Our guidelines are for four distinct practitioner groups: designers/developers, information security teams, cyber security experts, and C-suite employees. We offered guidelines for 14 different types of interventions within phishing education, training, and awareness, and for overcoming 19 different challenges that may arise in such interventions.
Our personalized guidelines aim to improve the effectiveness of current anti-phishing software development, deployment, and assessment practices. By reporting these guidelines to address the needs of anti-phishing practitioners, we aim to contribute to the ongoing efforts to mitigate the threat of phishing attacks. 





\section*{Acknowledgment}

The work has been supported by the Cyber Security Research Centre Limited whose activities are partially funded by the Australian Government’s Cooperative Research Centres Programme.

\bibliographystyle{ieeetr}

\end{multicols}
\end{document}